\documentclass[10pt, conference]{ieeeconf}      % Use this line for a4 paper

\IEEEoverridecommandlockouts                              % This command is only needed if 
\usepackage{graphicx}      % include this line if your document contains figures
\usepackage{amsmath}
\usepackage[utf8]{inputenc}
\usepackage{graphicx}
\usepackage{algorithmicx}
\usepackage{algorithm}
\usepackage{algpseudocode}
\usepackage{algorithm,algpseudocode}
\usepackage{amssymb}
\usepackage{float}
\usepackage{enumerate}
\usepackage{comment}
\usepackage{subfigure}
\usepackage[table,xcdraw]{xcolor}

\newtheorem{Problem}{\textbf{Problem}}

\title{\LARGE \bf
Optimization based coordination of autonomous vehicles in confined areas
}

\author{Stefan Kojchev$^{1}$, Robert Hult$^{2}$ and Jonas Fredriksson$^{3}$% <-this % stops a space
\thanks{*This work is partially funded by Sweden's innovation agency Vinnova, project number: 2018-02708.}% <-this % stops a space
\thanks{$^{1}$Stefan Kojchev is with Volvo Autonomous Solutions and the Mechatronics Group, Systems and Control, Chalmers University of Technology
        {\tt\small stefan.kojchev@volvo.com}; {\tt\small kojchev@chalmers.se}}%
\thanks{$^{2}$Robert Hult is with Volvo Autonomous Solutions, 41873 Göteborg, Sweden
        {\tt\small robert.hult@volvo.com}}%
\thanks{$^{3}$Jonas Fredriksson is with the Mechatronics Group, Systems and Control, Chalmers University of Technology, 41296 Göteborg, Sweden
        {\tt\small jonas.fredriksson@chalmers.se}}%
}

\newcommand{\tder}[2]{\frac{\mathrm{d}#1}{\mathrm{d}#2}}

\begin{document}

\maketitle
\thispagestyle{empty}
\pagestyle{empty}

%%%%%%%%%%%%%%%%%%%%%%%%%%%%%%%%%%%%%%%%%%%%%%%%%%%%%%%%%%%%%%%%%%%%%%%%%%%%%%%%
\begin{abstract}
Confined areas present an opportunity for early deployment of autonomous vehicles (AV) due to the absence of non-controlled traffic participants. In this paper, we present an approach for coordination of multiple AVs in confined sites. The method computes speed-profiles for the AVs such that collisions are avoided in cross-intersection and merge crossings. Specifically, this is done through the solution of an optimal control problem where the motion of all vehicles is optimized jointly. The order in which the vehicles pass the crossings is determined through the solution of a Mixed Integer Quadratic Program (MIQP). Through simulation results, we demonstrate the capability of the algorithm in terms of performance and satisfaction of collision avoidance constraints.
\end{abstract}

\section{Introduction}\label{Ch:Introduction}
It is believed that fully automated vehicles (AV) have the potential to drastically change the transport industry, both in terms of increased safety and efficiency \cite{b1}. The most drastic improvements are expected when a substantial part of the vehicles on public roads are fully automated, as in e.g., "robo-taxis" and hub-to-hub transports on highways. Unfortunately, managing the unpredictable conditions on public roads in a reliably safe manner has proven to be harder than initially expected, and the current state-of-the-art exhibits a lack of production-level maturity.

However, confined areas, such as mines, ports, and logistic centers, lack many of the difficult aspects of public road driving, and present use cases for near-future, large-scale deployment of automated vehicles. Within this context, the AVs form a component in transport solutions for commercial operations, where for example material-flow can be handled without human involvement. 

One of the challenges in such systems is the efficient coordination of multiple AVs use of mutually exclusive resources (MUTEX), such as intersections, narrow roads, work-stations (e.g. crushers, loading/unloading spots, etc.) and, in the case of electrified AVs, charging-stations. Poor coordination can lead to substantial decreases in productivity and energy-efficiency, reducing the benefits of automation.

The problem of handling mutual exclusive resources has been addressed for industrial robots \cite{b16}, \cite{b17}, where different scheduling algorithms have been explored. The coordination of multiple AVs, however, adds a different aspect to the challenge, where for example, the dynamics of the vehicles and the road topography play a significant part in the optimization problem. The coordination of automated vehicles at intersections has been widely discussed in the literature recently, see \cite{b2} for a comprehensive survey. In general, the problem is difficult to solve and has been formally shown to be NP-hard in \cite{b15}. Often relying on simplifying assumptions and heuristics, a number of methods has been presented that solve the problem  using, e.g.,  hybrid system theory \cite{b3}, reinforcement learning \cite{b4}, scheduling \cite{b5}, model predictive control (MPC) \cite{b6}, \cite{b7} or direct optimal control (DOC) \cite{b8}, \cite{b9}.

Coordination of AVs in confined areas has some distinct differences compared to the intersection scenarios often found in the literature. For instance, the full site-layout of confined areas is typically known at the planning stage, and it can often be expected that no non-controlled actors will disturb the execution of a plan once it's formed. The motion of each vehicle can therefore be planned from the start of a transport mission to its end. Planning for confined sites is thus benefited by methods that can handle long planning horizons. This is in contrast to the intersection coordination context found in the literature, where a cutout around the intersection proper is most often considered, with the vehicles arriving at speed \cite{b6}, \cite{b7}.
%Moreover, in confined areas, the motion plan must consider coordination of the vehicles' use of a number of resources besides intersections, e.g., different variety of collision zones, and the quality of the plan has a direct and significant impact on the profit margins of the transport solution provider.

In this paper, we formulate the MUTEX-coordination problem as an an optimal control problem. We adapt the two-stage heuristic procedure proposed in \cite{b9} to the confined-site context and employ it to solve the problem. Besides cross-intersections, we consider the merge-split MUTEX-zones, where the vehicles first join in on a common patch of road which after some distance separate. The approach is capable to optimize the vehicle trajectories over their full path and there are no limitations on the model that is used for the vehicles. Although the approach focuses on confined sites, the method of handling the mutual exclusion zones can be extendable to other scenarios as well (e.g., public road applications). 

From similar scenarios that have been considered, the authors in \cite{b10} and \cite{b11} propose an optimization approach for handling merge scenarios, which is a subset of the merge-split collision zone, and \cite{b12} uses a game-theoretical strategy for optimizing traffic flow through multiple intersection collision zones. The handling of multiple intersections and zones of different types were also identified in the survey \cite{b2} as topics for further work in this field.

The remainder of the paper is organized as follows: Section \ref{Ch:Problem formulation} formulates the problem that is solved in this paper. In Section \ref{Ch:Method} the method for solving the stated problem is presented, followed by Section \ref{Ch:Simulation results} where simulation results illustrate the coordination algorithm. Section \ref{Ch:Conclusions} concludes the work and provides some possible extensions.

\section{Problem formulation}\label{Ch:Problem formulation}

We consider $N_a$ fully automated vehicles on a road network with cross-intersection, path merges and path splits. The road network is assumed to be fully in a confined area, such that non-controlled traffic participants (e.g. manually operated vehicles, pedestrians, bicyclists etc.) are absent. We further assume that the paths of all vehicles, i.e., their routes through the road network are known, that no vehicle reverses, and that overtakes are prohibited.

\subsection{Vehicle modelling}
The motion of the vehicles along their path is described by
\begin{align}\label{eq:time_dynamics} 
\dot{p}_i(t) &= v_i(t) \\
\dot{x}_i(t) &= f_i(p_i(t), x_i(t),u_i(t)) \\ 
0 &\leq h_i(p_i(t),x_i(t),u_i(t)).
\end{align}
\noindent where $p_i(t) \in \mathbb{R}$ is the position, $x_i(t) \in \mathbb{R}^n$ the vehicle state,  $u_i(t)\in\mathbb{R}^m$ the control input, with $i\in \{1, \hdots, N_a \}$. The state is subdivided as  $x_i(t) = (v_i(t), z_i(t))$, with the speed along the path $v_i(t) \in \mathbb{R}$ and $z_i(t) \in \mathbb{R}^{n-1}$ collecting possible other states.  The functions $f_i$ and $h_i$, both assumed smooth, describes the dynamics and constraints that capture, e.g., actuator and speed limits respectively. 

\subsection{Vehicle modelling in the spatial domain}

For confined site optimization, it is beneficial to optimize the trajectories of the vehicles over their full paths. However, the time it takes a vehicle to traverse a path is dependent on the solution, and not known \textit{a-priori}. Consequently, it is inappropriate to plan the vehicle's motion with time as the independent variable. Due to this, the problem is reformulated in the spatial domain, using that $\frac{\mathrm{d}p_i}{\mathrm{d}t}= v_i(t)$ and $\mathrm{d}t = \mathrm{d}p_i/v_i(t)$. The formulation of the vehicle dynamics \eqref{eq:time_dynamics} in the spatial domain is
\begin{align}
    \tder{t_i}{p_i}&= \frac{1}{v_i(p_i)} \label{Eq: TimeDyn}\\
    \tder{x_i}{p_i} &= \frac{1}{v_i(p_i)}f_i(p_i,x_i(p_i),u_i(p_i)) \label{Eq: SpaceDyn}\\
    0 &\leq h(p_i,x_i,u_i) \label{Eq: State_Input_Const}.
\end{align}
\noindent where the position $p_i$ is the independent variable.

\subsection{Conflict zone modelling}

\begin{figure*}[htp]
    \centering
    \includegraphics[width=0.6\textwidth]{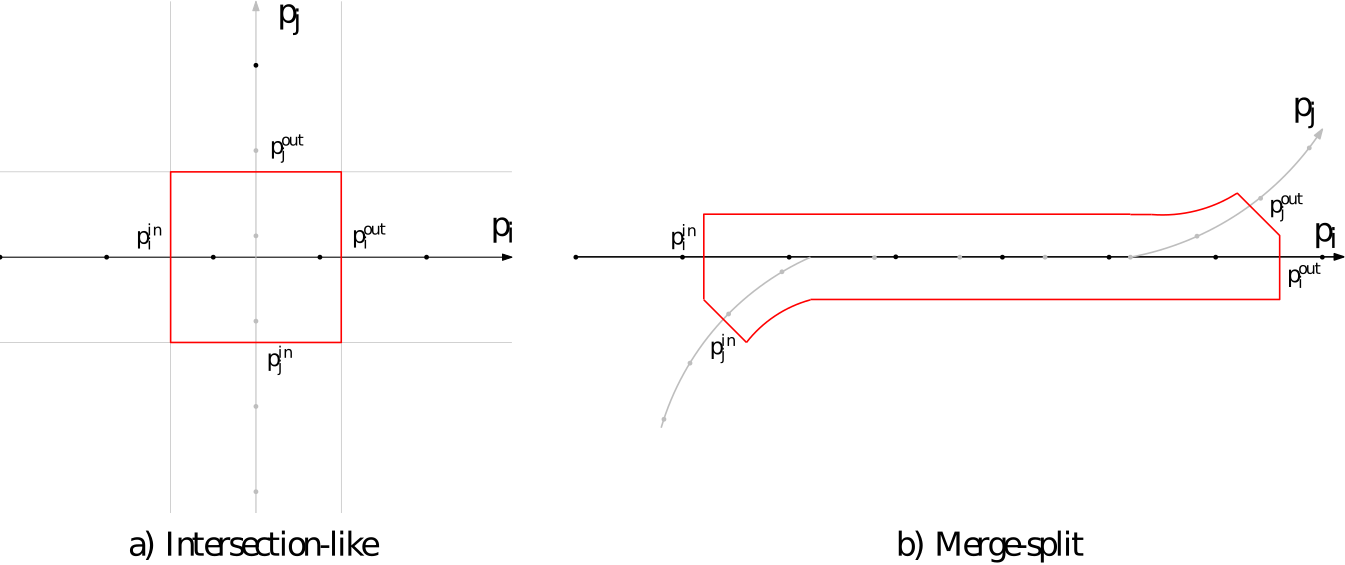}
    \caption{Types of conflict zones.}
    \label{fig:IntersectionsTogether}
\end{figure*}

A \textit{conflict zone} (CZ) is described by the entry and exit position $[p_i^{in},p_i^{out}]$ on the path of each vehicle. From the known positions, the time of entry and exit of vehicle $i$ is $t_i^{in}=t_i(p_i^{in})$, $t_i^{out}=t_i(p_i^{out})$, respectively.
In this paper, two types of conflict zones are considered, as depicted in Figure \ref{fig:IntersectionsTogether}: the ``intersection-like" and the ``merge-split". We let $\mathcal{I} = \left \{ I_1, I_2,...,I_{r_0} \right \}$ denote the set of all intersections in the confined site, with $r_0$ being the total number of intersection CZs,  and let $\mathcal{Q}_r = \left \{ q_{r,1}, q_{r,2},...,q_{r,l} \right \}$ denote the set of vehicles that cross an intersection $I_r$. In the intersection-like CZ, it is desired to only have one vehicle inside the CZ, i.e., not allowing the vehicle $j$ to enter the CZ before vehicle $i \neq j$ exits the CZ, or vice-versa. The order in which the vehicles cross the intersection $I_r$ is denoted $\mathcal{O}^{\mathcal{I}}_r = \left ( s_{r,1}, s_{r,2},...,s_{r,|\mathcal{Q}_r|} \right )$, where $s_{r,1}, s_{r,2},...$ are vehicle indices and we let $\mathcal{O}^\mathcal{I}=\left \{\mathcal{O}^\mathcal{I}_1, \hdots, \mathcal{O}^\mathcal{I}_r\right \}$. A sufficient condition for collision avoidance for the $r$-th intersection CZ can be formulated as
%\begin{equation}\label{Eq: IntersectionConst}
%    t_i(p_i^{out})\leq t_j(p_j^{in}) \; \; \textup{\textbf{OR}} \; \; t_j(p_j^{out})\leq t_i(p_i^{in}) \;\; \forall i,j \; i \neq j.
%\end{equation}
%\begin{align}\label{Eq: IntersectionConst}
%    t_{s_{r,i}}(p_{s_{r,i}}^{out})\leq t_{s_{r,i+1}}(p_{s_{r,i+1}}^{in}), \; &i \in \mathbb{I}_{\left [ 1, |Q_r|-1 \right ]}, \\ \nonumber
%    &r \in \mathbb{I}_{\left [ 1, |I_r|-1 \right ]}, I_r \in \mathcal{I}.
%\end{align}
\begin{align}\label{Eq: IntersectionConst}
    t_{s_{r,i}}(p_{s_{r,i}}^{out})\leq t_{s_{r,i+1}}(p_{s_{r,i+1}}^{in}), \; &i \in \mathbb{I}_{\left [ 1, |\mathcal{Q}_r|-1 \right ]},
\end{align}
\noindent where $t$ is determined from \eqref{Eq: TimeDyn}.
In the merge-split CZ case, let $\mathcal{M} = \left \{ M_1, M_2,...,M_{w_0}  \right \}$ denote a set of all merge-split zones, with $w_0 $ being the total number of merge-split CZs in the site and $\mathcal{Z}_w = \left \{ z_{w,1}, z_{w,2},...,z_{w,h} \right \}$ denote the set of vehicles that cross the merge-split CZ $M_w$. For efficiency, it is desirable to have several vehicles in the zone at the same time, instead of blocking the whole zone. This requires having rear-end collision constraints once the vehicles have entered the CZ safely. In this case, the order in which the vehicles enter the zone is denoted as $\mathcal{O}^{\mathcal{M}}_w = \left ( s_{w,1}, s_{w,2},...,s_{w,|\mathcal{Z}_w|}\right )$, and we let $\mathcal{O}^\mathcal{M}=\left \{\mathcal{O}^\mathcal{M}_1, \hdots, \mathcal{O}^\mathcal{M}_w\right \}$. The collision avoidance requirement for this $w$-th CZ is described with the following constraints: 
\begin{subequations}
\label{Eq:MergeSplit_ORConst}
\begin{align}\label{Eq: MS_In}
t_{s_{w,i}}(p_{s_{w,i}}^{in}) + \Delta t &\leq t_{s_{w,i+1}}(p_{s_{w,i+1}}^{in}+c) \\ \nonumber
t_{{s_{w,i}},k_i} + \Delta t &\leq t_{s_{w,i+1}}(p_{{s_{w,i}},k_i} - p_{s_{w,i}}^{in} + p_{s_{w,i+1}}^{in} + c), \\  & \;\;\;\;k_{s_{w,i}}^{in} \leq k_i \leq k_{s_{w,i}}^{out} \label{Eq: MS_Middle}  \\ 
t_{s_{w,i}}(p_{s_{w,i}}^{out}) + \Delta t &\leq t_{s_{w,i+1}}(p_{s_{w,i+1}}^{out}+c), \label{Eq: MS_Out}\\ \nonumber
&i \in \mathbb{I}_{\left [ 1, |\mathcal{Z}_w|-1 \right ]}.
\end{align}
\end{subequations}
That is, while in the CZ, the vehicles must be separated by at least a time-period $\Delta t_i$ and a distance $c_i$ or by $\Delta t_j$ and $c_j$, depending on if vehicle $j$ is in front of vehicle $i$ or vice versa.
%is added for an additional constant time gap between the vehicles. However, only including a time gap would lead to the problem that at low speeds the vehicles would be very close to each other, unless $\Delta t$ is not very large. Large $\Delta t$, on the other hand, would lead to too large distance between the vehicles at high speeds. Therefore, a constant $c$ is included as well.
%As we assume that the vehicles do not reverse, the velocity is at all times larger than zero, meaning that $c>0$. 
This is equivalent to the standard offset and time-headway formulation often used in automotive adaptive cruise controllers.

%Introduce zone sets here, one for intersections, one for merge and appropriate sets for involved vehicles.

%The total number of CZ in the scenario is denoted as $N_{CZ} = \textup{card}(\mathcal{I}) + \textup{card}(\mathcal{M})$ and the crossing order for all the collision zones is denoted as $\mathcal{O} = \mathcal{O}_r\cup\mathcal{O}_w$. One can easily conclude that the combinatorial problem quickly grows with the increase of vehicles and conflict zones.

\subsection{Discretization}
The independent variable is discretized as $p_i = (p_{i,1}, \hdots, p_{i,N_i})$, where $p_{i,N_i}$ indicates the end position for vehicle $i$, and the input is approximated using zero order hold such that $u(p) = u_{i,k}, p\in[p_{i,k}, p_{i,k+1}[$. The equations (\ref{Eq: TimeDyn}), (\ref{Eq: SpaceDyn}) are (numerically) integrated on this grid, giving the ``discretized" state transition relation 
\begin{equation}\label{Eq: SystemDynamics}
    \begin{bmatrix}
    t_{i,k+1} \\
    x_{i,k+1}
    \end{bmatrix} = F(x_{i,k},u_{i,k},p_{i,k},p_{i,k+1}) 
\end{equation}
where $F$ denotes the integration of (\ref{Eq: TimeDyn}), (\ref{Eq: SpaceDyn}) from $p_{i,k}$ to $p_{i,k+1}$.

\subsection{Optimal coordination problem}

The problem of finding the optimal vehicle trajectories that avoid collisions can be formalized as: 

\Problem (Optimal coordination problem) Obtain the optimal state and control trajectories $\mathcal{X}^{*} = \left \{x_1^*,...,x_{N_a}^*\right \}$, $\mathcal{U}^{*} = \left \{u_1^*,...,u_{N_a}^*\right \}$, given the initial state $\mathcal{X}_{0} = \left \{x_{1,0},...,x_{N_a,0}\right \}$, by solving the optimization problem 
\begin{subequations}
\label{Eq: Main_OCP}
\begin{align} 
        \underset{x_{i,k},u_{i,k}, \mathcal{O}^\mathcal{I}, \mathcal{O}^\mathcal{M}}{\textup{min}} \;\; &\sum_{i = 1}^{N_a}J_i\left ( x_{i,k},u_{i,k} \right )  \\
\textup{s.t} \;\;\; & \textup{initial states} \; \; x_{i,0} = \hat{x}_{i,0}, \forall i \label{Eq: InitialGuess}  \\
&\textup{system dynamics}\; \; (\ref{Eq: SystemDynamics}), \; \forall i, \forall k  \label{Eq: OCPSysDyn} \\
&\textup{state and input constraints}  \; \; (\ref{Eq: State_Input_Const}), \; \forall i, \forall k \label{Eq: OCPStateInpConst} \\
& \textup{safety constraints} \; \;  (\ref{Eq: IntersectionConst}), (\ref{Eq:MergeSplit_ORConst}), \; \forall i, \forall k \label{Eq: OCPSafetyCosnt}
\end{align}
\end{subequations}
Note in particular that this involves finding the crossing orders $\mathcal{O}^\mathcal{I}$, $\mathcal{O}^\mathcal{M}$, which makes the problem combinatorial and difficult to solve.

\section{Method}\label{Ch:Method}

Problem 1 can be stated as a Mixed Integer Nonlinear Program (MINLP), where the crossing order correspond to the ``integer part" and the state and control trajectories corresponds to the ``NLP part". However, finding a solution to MINLP problems is known to be difficult, especially when the constraints or the objective function are non-convex \cite{b13}. Therefore, a common procedure is to apply an approach where the integer part of the solution is obtained first using a heuristic, and the continuous part of the solution thereafter is obtained by solving the nonlinear program (NLP) that results from fixing the integers to the values found with the heuristic. In this paper, we follow this approach. The heuristic that is used is similar to that of \cite{b9} which approximates the integer part of the solution of (\ref{Eq: Main_OCP}) by solving a Mixed Integer Quadratic Problem (MIQP). With the integer solution given, the state and control trajectories are obtained by solving the ``fixed-order coordination" NLP, i.e., Problem (\ref{Eq: Main_OCP}) with fixed crossing orders $\mathcal{O}^\mathcal{I}$, $\mathcal{O}^\mathcal{M}$.

\subsection{Crossing order heuristic}
The crossing order heuristic follows the procedure proposed in \cite{b9} where the coordination problem is approximated as time-slot scheduling MINLP in $\mathbf{T}$, where $\mathbf{T} = (T_i,...,T_{N_a})$ and $T_i$ is the set of entry and exit times for vehicle $i$ for all the CZ it encounters. With this formulation, the integer part of the solution, i.e., the crossing orders, are treated in $\mathbf{T}$ space. The problem can be formulated as
\begin{subequations}
\label{Eq: MINLP}
\begin{align}
    \underset{\mathbf{T}, \mathcal{O}^\mathcal{I}, \mathcal{O}^\mathcal{M}}{\textup{min}} \;\; &\sum_{i=1}^{N_a}V_i(T_i) \\
    \textup{s.t.} \;\; & T_i\in \mathrm{dom}(V_i) \ \forall i, 1, \hdots, N_a, (\ref{Eq: IntersectionConst}), (\ref{Eq: MS_In}), \eqref{Eq: MS_Out},
\end{align}
\end{subequations}
where $V_i$ is the value function and $\mathrm{dom}(V_i)$ is the domain of it. The definition of $V_i(T_i)$ follows next.

Note that the solution to \eqref{Eq: MINLP} is an approximation of the solution to \eqref{Eq: Main_OCP}. In particular, the rear-end collision avoidance constraints \eqref{Eq: MS_Middle} are not enforced and could therefore be violated.
\subsubsection{The vehicle problem}
The function $V_i(T_i)$ is the optimal value function of the following parametric optimal control problem, denoted as the ``Vehicle Problem":
\begin{subequations}
\label{Eq: ParametricNLP}
\begin{align}
V_i(T_i) = \;\; &\underset{x_i,u_i}{\textup{min}} \;\; J_i(x_i,u_i) \\ 
&\textup{s.t.} \;\; (\ref{Eq: InitialGuess}),(\ref{Eq: SystemDynamics}),(\ref{Eq: State_Input_Const}) \\
& \hspace{0.7cm} \xi _i^{in} = t_i(p_i^{in}) \\
& \hspace{0.7cm} \xi _i^{out} = t_i(p_i^{out}), 
\end{align}
\end{subequations}
\noindent with $\xi _i$ being the parameter over which we optimize i.e. the in- and out-times of the CZs. Note that $\mathrm{dom}(V_i)$ thereby is the set of parameters $T_i$ for which \eqref{Eq: ParametricNLP} has a solution.

\subsubsection{An MIQP-based heuristic}
The time-slot scheduling problem \eqref{Eq: MINLP} is still a difficult problem, with non-convex objective function and constraints. Motivated by the availability of efficient solvers of MIQPs, we therefore consider the following approximation of \eqref{Eq: MINLP} using the first and second order Taylor expansion of the objective function and constraints: 
\begin{subequations}
\label{Eq: MIQP}
\begin{align}
    \underset{\mathbf{T}}{\textup{min}} \;\; &\sum_{i=1}^{N_a}\frac{1}{2}T_i^T\triangledown ^2V_iT_i + \left ( \triangledown V_i-\triangledown^2V_iT_i^{[0]} \right )^TT_i \\
    \textup{s.t.} \;\; & (\ref{Eq: IntersectionConst}), (\ref{Eq: MS_In}), \eqref{Eq: MS_Out},
\end{align}
\end{subequations}
\noindent where the derivatives are evaluated at $T_i^{[0]}$. Note that besides that the problem now is an MIQP, the constraint $\mathrm{dom}(V_i)$ is removed. As discussed in \cite{b9}, the boundary of this constraint are all timeslots where the control is fully saturated prior to the intersection. This typically occur only in scenarios with a large number of vehicles that are forced to resolve conflicts over short distances, at high speeds. As such situations can be avoided in the target application, the constraint is removed.

%This is motivated by the conclusions in \cite{b9}, where the feasibility of the heuristic and solution quality is also discussed. 

The solution to the MIQP problem provides an approximately optimal time-slot schedule $\mathbf{T}$ and therefore approximate crossing orders $\hat{\mathcal{O}}^\mathcal{I}$, $\hat{\mathcal{O}}^\mathcal{M}$. 

\paragraph*{A comment on differentiation of $V_i(T_i)$}
From the solution of the parametric NLP \eqref{Eq: ParametricNLP}, the first and second order derivatives w.r.t. the parameter can be calculated using parametric sensitivity analysis \cite{b14}. In particular, this can be done at a low additional cost when the optimal problem is solved. Note that for a solution to problem (\ref{Eq: ParametricNLP}) feasible entry and exit times, i.e., feasible values for the parameters, for each CZ that vehicle $i$ encounters are necessary. These times can, for example, be computed by solving the optimization problem (\ref{Eq: Main_OCP}) without safety constraints (\ref{Eq: IntersectionConst}), or a forward simulation of the vehicles with, for example, an LQR controller. 

\subsection{Fixed-order NLP}

\begin{figure*}
    \centering
    \subfigure[]{\includegraphics[width=0.28\textwidth]{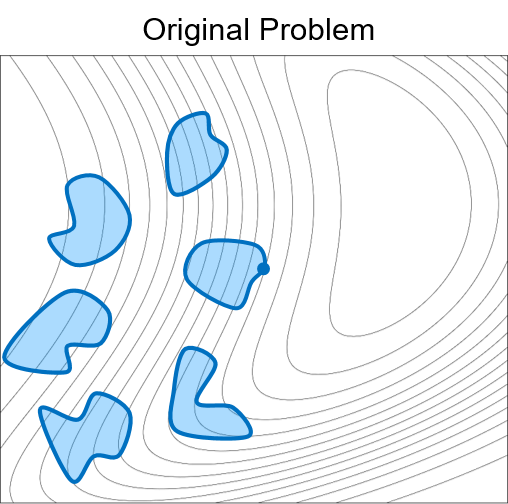}} 
    \subfigure[]{\includegraphics[width=0.28\textwidth]{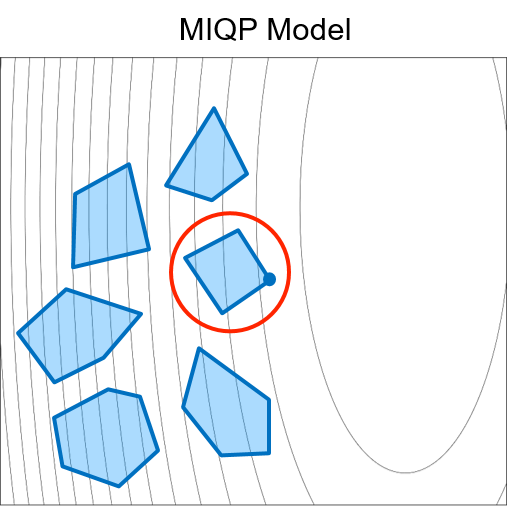}} 
    \subfigure[]{\includegraphics[width=0.28\textwidth]{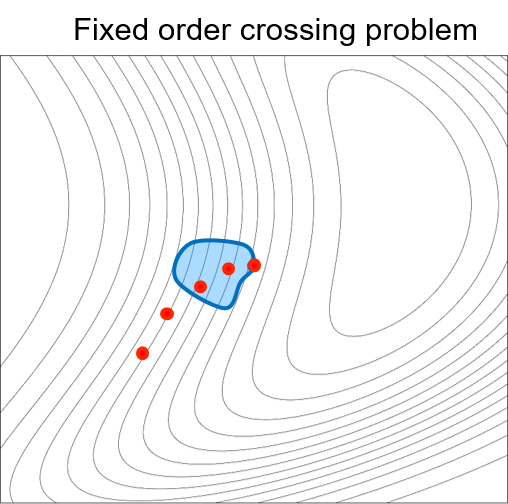}}
    \caption{Illustration of the heuristic approach. As seen, in (a) the original problem is a non-convex problem with a disconnected solution space (each compact subset corresponding to an integer combination), whose solution is, typically, difficult to obtain. The MIQP approximation (b) of the original problem allows for finding a region where the solution of (a) could be. Finally, through the fixed-order NLP (c), the approximate optimal solution is acquired.}
    \label{fig: HeuristicIllustration}
\end{figure*}

With the found crossing order, the integer part of the solution is obtained, making \eqref{Eq: Main_OCP} an NLP. Obtaining the optimal state and control trajectories is thus found through solving the fixed-order coordination problem 
\begin{subequations}
\label{Eq: FixedOrderNLP}
\begin{align}
    \underset{x_{i,k},u_{i,k}}{\textup{min}} \;\; &\sum_{i = 1}^{N_a}J_i\left ( x_{i,k},u_{i,k} \right )  \\
    \textup{s.t} \;\;\; & (\ref{Eq: InitialGuess})-(\ref{Eq: OCPSafetyCosnt}), \; \forall i, \forall k \\
    &\mathcal{O}^\mathcal{I} = \hat{\mathcal{O}}^\mathcal{I}, \;\; \mathcal{O}^\mathcal{M}=\hat{\mathcal{O}}^\mathcal{M}
\end{align}
\end{subequations}
The two stage approximation approach is summarized in Algorithm \ref{Al: ApproximationAlgrthm} and depicted in Figure \ref{fig: HeuristicIllustration}.
\begin{algorithm}
\caption{Two stage approximation algorithm}
\begin{flushleft}
        \textbf{Input:} $N_a,\mathcal{I}, \mathcal{Q}_r, \mathcal{M}, \mathcal{Z}_w$, vehicle paths\\
        \textbf{Output:} $\mathcal{X}^{*}, \; \mathcal{U}^{*}$
\end{flushleft}
\begin{algorithmic}[1] 
\State $\forall i$: Obtain feasible entry and exit times $T_i^{[0]}$ by, e.g., solving NLP (\ref{Eq: Main_OCP}) w/o safety constraints (\ref{Eq: IntersectionConst}). \label{Al: ApproximationAlgrthm}
\State $\forall i$: Solve NLP (\ref{Eq: ParametricNLP}) and calculate the first and second order sensitivities ($\triangledown V_i, \triangledown ^2 V_i$).
\State Solve MIQP (\ref{Eq: MIQP}) for the approximate crossing orders $\hat{\mathcal{O}}^\mathcal{I}$, $\hat{\mathcal{O}}^\mathcal{M}$.
\State Solve the fixed-order NLP (\ref{Eq: FixedOrderNLP})  using $\hat{\mathcal{O}}^\mathcal{I}$, $\hat{\mathcal{O}}^\mathcal{M}$ and obtain $\mathcal{X}^{*}, \; \mathcal{U}^{*}$. 
\end{algorithmic}
\end{algorithm}

\section{Simulation results}\label{Ch:Simulation results}
In this section, we present a simulation example showing the operation and performance of the coordination algorithm.
\subsection{Simulation setup}
The vehicles are modelled as a triple integrator $\dddot{x} = u$, whereby the spatial model is: 
\begin{equation} 
    \begin{bmatrix}
    \tder{t_i}{p_i} \\
    \tder{v_i}{p_i} \\
    \tder{a_i}{p_i}
    \end{bmatrix} = \begin{matrix}
\frac{1}{v_i}\\ 
\frac{a_i}{v_i}\\ 
\frac{u_i}{v_i}
\end{matrix},
\end{equation}
\noindent where $a_i$ is the acceleration and $u_i$ is the jerk. The dynamics are discretized using multiple shooting (with $N$ shooting points) and an Explicit Runge-Kutta-4 (ERK4).

The constraints are chosen as bounds on the speed and longitudinal acceleration $(\underline{v}_i \leq v_{i,k} \leq \overline{v}_i, \; a_{i,k} \leq \overline{a}_{i,lon})$ in order to obey speed limits and physical constraints. The lower bound on the speed is necessary to be non-zero in order to avoid singularity in the vehicle model. The vehicles are expected to manoeuvre on curved roads, thus, it is needed to consider the lateral forces they experience. As the one-dimensional model that is used in this paper does not account for lateral motion, the following constraint is enforced, as similarly proposed in \cite{b8}.
\begin{equation}\label{Eq: StateConst2}
    \left ( \frac{a_{i,k}}{a_{i,lon}} \right )^2 + \left ( \frac{\kappa_i(p_{i,k})v_{i,k}}{\overline{a}_{i,lat}} \right )^2 \leq 1,
\end{equation}
\noindent where $\overline{a}_{i,lat}$ is the lateral acceleration limit and $\kappa_i(p_{i,k})$ is the road curvature, that is assumed to be available at every point along the path.  

While other objectives could be used, we consider the following trade-off between minimization of the total travel time and the squares of the longitudinal acceleration and longitudinal jerk:
\begin{align}
&J(x_i(p_i),u_i(p_i)) = \\ \nonumber
&\int_{p_{i,0}}^{p_{i,N_i}}  \left ( P_ia_i(p_i)^2 + Q_ij_i(p_i)^2 \right )\frac{1}{v_i(p_i)}\mathrm{d}p_i + R_it_i(p_{i,N_i}),
\end{align}
\noindent where $P_i,Q_i,R_i$ are the appropriate weights. This objective is integrated with Forward Euler, leading to the ``discretized" objective expression
%, it is of interest to minimize the change of acceleration and jerk in order for the vehicles to operate in a comfortable way. Minimizing the acceleration and jerk would also lead to a reduction in fuel/energy consumption. This requirement leads to the following objective function for each vehicle:
\begin{align}
    &J_i\left ( x_{i,k},u_{i,k} \right ) = \\ \nonumber
    &\sum_{k=1}^{N}\left (\left ( P_ia_{i,k}^2 + Q_ij_{i,k}^2 \right ) \frac{\Delta p_{i,k}}{v_{i,k}}\right ) + R_it_{i,N},
\end{align}
\noindent with $\Delta p_{i,k} = p_{i,k+1}-p_{i,k}$.

\begin{figure}[h]
    \centering
    \includegraphics[width=0.33\textwidth,height=7.5cm]{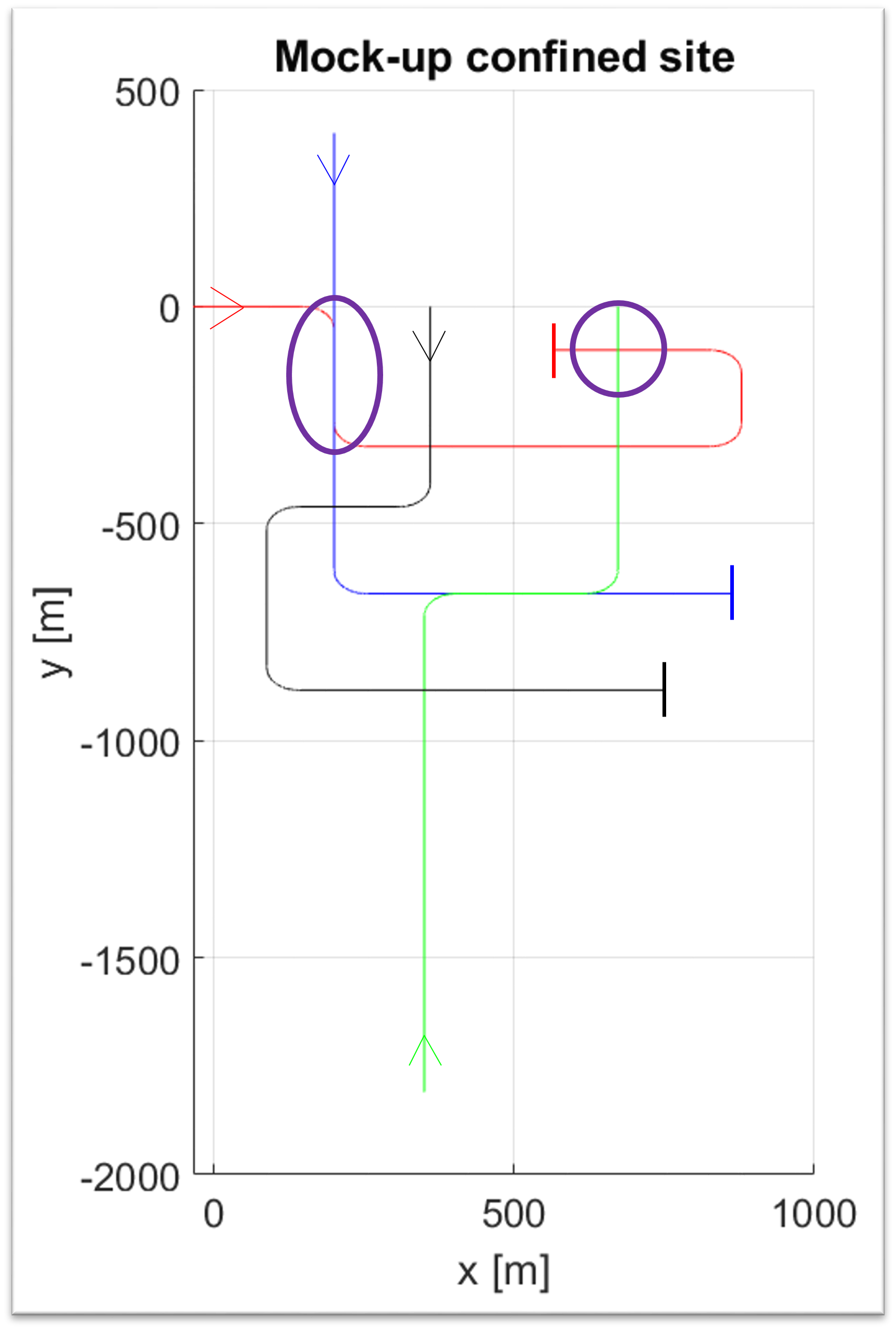}
    \caption{Mock-up confined site area. The lines indicate the vehicle's path through the site with different color per vehicle. The arrow is the direction of movement for the vehicles and the solid line is indicates the path end. The CZs that are investigated in the simulation scenario are circled over in this figure.}
    \label{Fig: SiteMap}
\end{figure}

For evaluation purposes, we are investigating a mock-up confined area with four vehicles as depicted in Figure \ref{Fig: SiteMap}. For this scenario, the vehicle paths are permanent throughout the simulation. There are in total two merge-split CZs and five intersection CZs. Every vehicle starts from an initial velocity of 50 $\textup{km/h}$ and the chosen bounds are $\underline{v}_i = 3.6 \; \textup{km/h}$, $\overline{v}_i = 90 \; \textup{km/h}$, $\overline{a}_{i,lon} = 4 \; [\textup{m}/\textup{s}^2]$, $\overline{a}_{i,lat} = 2 \; [\textup{m}/\textup{s}^2]$. The weight coefficients for the optimal problem are: $P_i = 1, \; Q_i = 1, \; R_i = 10$, and are a suitable trade-off between the objectives for this scenario. The total shooting points for this scenario is $N = 100$ and is same for all vehicles. In the intersection-like scenario the CZ is created with 5 meter margin ahead of and behind the collision point, where as in the merge-split scenario the margin is 15 meters for both the entry and exit collision point. The margin should account for the length of a vehicle. For the merge-split CZ it is also desired to keep at least 0.5s margin between the vehicles. 
We utilize the CasADi \cite{b18} toolkit with the IPOPT \cite{b19} solver for solving the optimization problems (\ref{Eq: ParametricNLP}), (\ref{Eq: FixedOrderNLP}) and Gurobi for solving the MIQP problem (\ref{Eq: MIQP}).
\subsection{Discussion of results}
In the following, the results for this simulation scenario are presented. In particular, we demonstrate the uncoordinated and coordinated results for one merge-split zone and one intersection zone, both circled over in Figure \ref{Fig: SiteMap}. The uncoordinated results are obtained as the individual optimum of each vehicle, i.e., the speed profiles are obtained by solving the optimization problem without any safety constraints. 
\begin{figure*}
    \centering
    \begin{minipage}{0.4\textwidth}
        \centering
        \includegraphics[width=\textwidth,height=5.5cm]{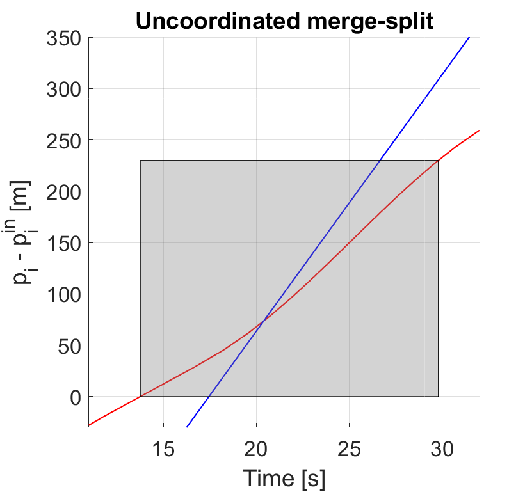} 
        \caption{Uncoordinated crossing for the merge-split CZ for the red and blue vehicle. The gray box is the occupancy time in the zone for the red vehicle. While it is allowed for the blue vehicle to enter the CZ whilst the red vehicle is in it, an intersection between the vehicles translates to a collision between the vehicles.}
        \label{Fig: UncoordinatedMergeSplit}
    \end{minipage}\hfill
    \begin{minipage}{0.4\textwidth} 
        \centering
        \includegraphics[width=\textwidth,height=5.5cm]{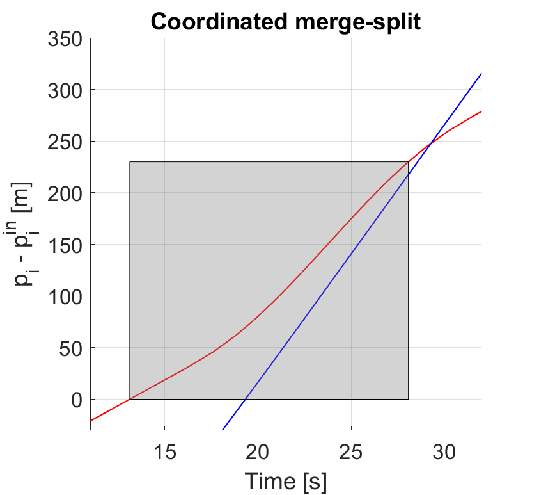}
        \caption{Coordinated crossing for the merge-split CZ for the red and blue vehicle. The gray box is the occupancy time in the zone for the red vehicle. The blue vehicle enters the CZ whilst the red vehicle is in and keeps the desired time gap.}
        \label{Fig: CoordinatedMergeSplit}
    \end{minipage}
\end{figure*}
\begin{figure}
    \centering
    \includegraphics[width=0.4\textwidth,height=7.0cm,keepaspectratio]{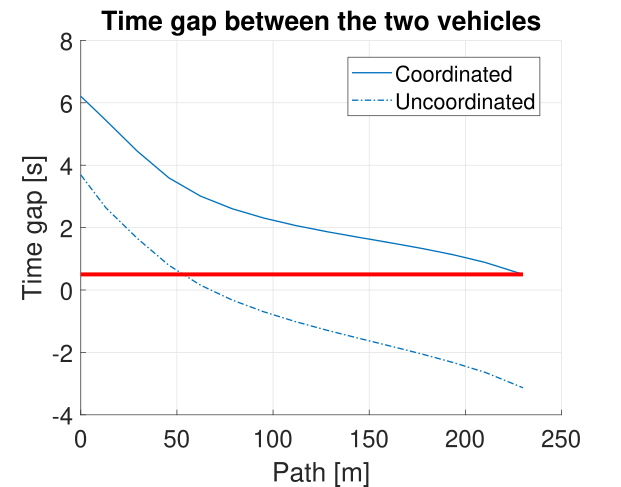}
    \caption{The time gap between the red and blue vehicles in the merge-split CZ for the coordinated and uncoordinated case. The bold red line represents the minimum time gap that is desired to be kept between the vehicles. Crossing this border results in a collision between the vehicles.}
    \label{Fig: TimeGap_MS}
\end{figure}
\begin{comment}
\begin{figure*}
    \centering
    \begin{minipage}{0.4\textwidth}
        \centering
        \includegraphics[width=\textwidth,height=7cm]{UncoordinatedIntersection.eps} 
        \caption{Uncoordinated crossing for an intersection CZ for the red and green vehicle.}
        \label{Fig: UncoordinatedIntersection}
    \end{minipage}\hfill
    \begin{minipage}{0.48\textwidth}
        \centering
        \includegraphics[width=\textwidth,height=7cm]{CoordinatedIntersection.eps}
        \caption{Coordinated crossing for an intersection CZ for the red and green vehicle.}
        \label{Fig: CoordinatedIntersection}
    \end{minipage}
\end{figure*}
\end{comment}
Figure \ref{Fig: UncoordinatedMergeSplit} and Figure \ref{Fig: CoordinatedMergeSplit} illustrate the uncoordinated and coordinated crossing of the merge-split CZ, respectively. In the figures, the path trajectories of the involved vehicles are depicted. The path trajectories have an offset of their entry position to the merge-split CZ, meaning that, for both vehicles, a position of zero indicates the entry position to the CZ. The gray rectangle depicts the CZ, with entry and exit times for the red vehicle, since that is the first vehicle that enters the zone in both cases. As mentioned, in this type of CZ it is desirable to have multiple vehicles inside the zone whilst keeping a minimum gap between them. This means that while it is allowed for both trajectories to be inside the zone, it is not allowed for the trajectories to intersect whilst in the zone. In the uncoordinated case, it can be noticed that the blue vehicle enters the CZ later than the red vehicle, however, intersects the red vehicle and exits the zone first. The interpretation of this behaviour is that inside the CZ the blue vehicle ``overruns" the red vehicle, meaning that a collision occurs. In the coordinated case, the vehicles are capable of being inside the CZ at the same time and keeping at least a minimum designated gap between each other. Note that an intersection of the trajectories outside the CZ is not relevant as they are no longer on a shared road. By allowing the vehicles to be both in the CZ, and not blocking the whole zone for one vehicle, the throughput is increased meaning that the vehicles could get to their end destination earlier. This could translate to increased efficiency or productivity of the site. To supplement the figures, Figure \ref{Fig: TimeGap_MS} depicts the time gap between the vehicles while they are in the CZ. In the figure, the moment where the blue vehicle ``overruns" the red vehicle in the uncoordinated case can be noticed, as well as the fact, that the desired minimum gap is kept for the coordinated case. 

\begin{figure}
    \centering
    \includegraphics[width=0.45\textwidth,height=8.0cm,keepaspectratio]{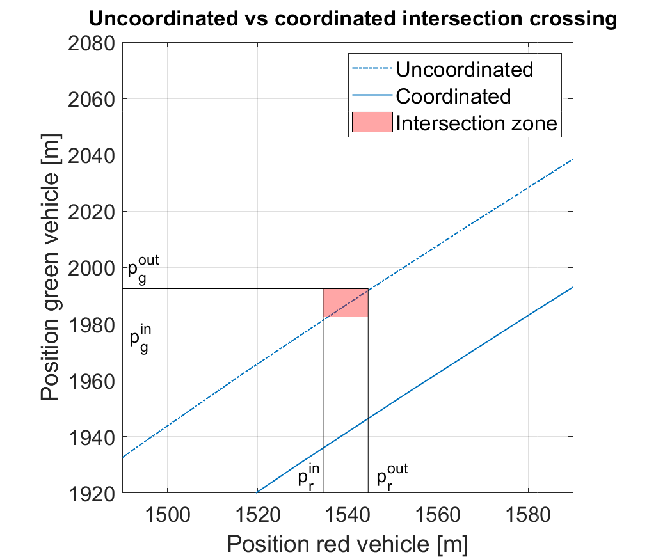}
    \caption{Uncoordinated and coordinated crossing of the intersection zone. An intersection of the depicted trajectory with the intersection zone is equivalent to both vehicles being in the CZ at the same time.}
    \label{Fig: IntersectionCrossing}
\end{figure}

\begin{figure}
    \centering
    \includegraphics[width=0.45\textwidth,height=9cm,keepaspectratio]{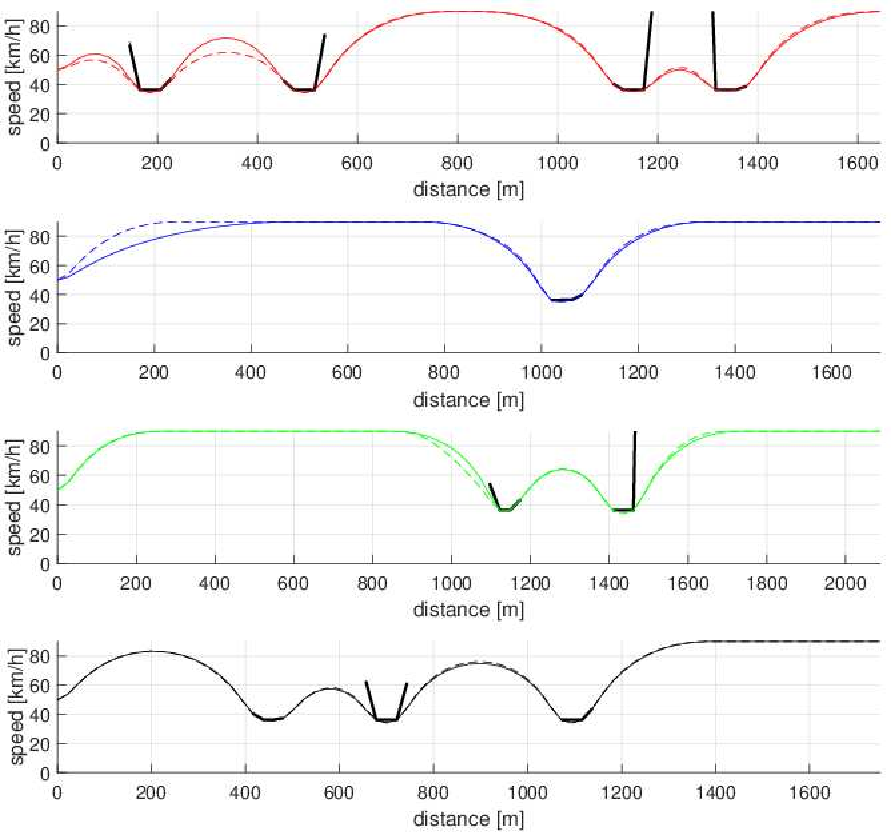}
    \caption{Speed profiles for the uncoordinated and coordinated vehicles. The uncoordinated speed profiles are depicted with dashed lines, the coordinated speed profiles are illustrated with full lines and the curvature constraints based on \eqref{Eq: StateConst2} are depicted with bold black lines.}
    \label{fig:VehicleSpeed}
\end{figure}

In the scenarios where two (or more) vehicles intersect, it is desired to only have one vehicle in the zone. The reasoning for this decision is that the CZ in this case occupies a relatively small patch of road and blocking off the whole zone for one vehicle will not result in a major loss of throughput efficiency. Figure \ref{Fig: IntersectionCrossing} depicts the uncoordinated and coordinated vehicle trajectories along with the intersection zone. For this dependency, a collision is defined if the trajectory intersects the intersection zone, which is equivalent to the two vehicles being in the zone at the same time. It is noticeable that in the uncoordinated case the intersection zone is intersected, whilst the coordinated vehicles satisfy the collision constraints as they successfully avoid being in the CZ at the same time.  
Figure \ref{fig:VehicleSpeed} depicts the vehicle's speed profile throughout the site for both the uncoordinated and coordinated results. As the objective function is to minimize the acceleration and jerk, it is noticeable that the speed profiles are smooth throughout the whole site and would not be challenging to follow in an application case. In the speed plots, the black lines represent the curvature constraints based on \eqref{Eq: StateConst2}. Since the algorithm considers all conflict zones between the vehicles from the beginning till the end of their path, undesired occupancy of the zone (i.e., collisions) can be avoided with minor changes of the speed profile over the whole route. In contrast, coordinating the vehicles in a cut-out around the CZ would require larger changes in speed in order to avoid conflicts. 
It is worth mentioning that in the uncoordinated case for the complete site, two more collisions occur. One in the other merge-split CZ between the blue and green vehicle and another in the intersection CZ of the blue and black vehicle. When the vehicles are coordinated, these collisions are avoided. For sake of brevity, we refrain from depicting these results as the critical behaviour is similar to the above-mentioned collision zones. 

The simulation scenario is implemented in MATLAB on a 2.90GHz Intel Xeon computer with 32GB of RAM. The total solver computational time for solving the example is 1.815 seconds, of which the MIQP requires 0.08 seconds and the fixed-order NLP requires 0.709 seconds and the remainder is the required time to solve the individual vehicle problems. The computation time for the MIQP scales exponentially with the number of decision variables, while the fixed-order NLP scales cubicly with the number of vehicles.

\section{Conclusions and future work}\label{Ch:Conclusions}

In this paper, we have presented an optimal control based approach for coordination of automated vehicles for sites where multiple collision zones of different types could occur. The approach optimizes the trajectories of the vehicles for their whole path length, taking all collision zones into account. To compute the crossing order for the collision zones, the approach relies on an MIQP-based heuristic. Simulation results demonstrate the capability to coordinate the vehicles without collisions. In future work, we intend to investigate improvement on the solver computation time, closed-loop behaviour as well as including higher fidelity vehicle models.

\end{document}